# Fluorescence for high school students

Niek G Schultheiss[1,2] and Tom W Kool[3]

[1]Nikhef, Amsterdam, the Netherlands

[2]Zaanlands Lyceum, Zaandam, the Netherlands

[3]Van 't Hoff Institute for Molecular Sciences, University of Amsterdam
the Netherlands

## Abstract

In a not obligatory series of lessons for high school students in the Netherlands we discuss the fluorescence aspects of anthracene. These lessons were developed because HiSPARC (High school Project on Astrophysics Research with Cosmics) detection of cosmic rays are available for different secondary schools. With the help of special designed scintillator detection stations, containing anthracene, cosmic rays can be detected.

Fluorescence of anthracene is one of the topics discussed in these series of extra curricular lessons aimed at excellent pupils working on cosmic radiation within the HiSPARC - project.

## Introduction

One of us (N.S.) developed a so-called Route-Net for secondary school students [1]. This Route-Net consists of different physics modules, each takes about two hours intensive study time. From the age of about sixteen the student has the opportunity to study different aspects of modern physics which are not taught in the 'normal' curriculum of the physics and chemistry program. After finishing a certain route the student is able to write a 'thesis' (piece of work) in the last year of the secondary school in the Netherlands. This will take about eighty hours study time. For the practical part of this thesis the student can measure cosmic rays, which is possible with the help of detection stations (in the form of ski-boxes) situated at the roof of the school. Detection stations of this type are not only located in the Netherlands, but also in the U.K. as well as Denmark. Fluorescence is one of the steppingstones in Route-Net. These modules form a grid of interconnecting downloadable hand-outs and are taught under the guidance of a teacher. The steppingstones start from the second grade of secondary education and can be finished with a 'thesis'. For the practical part of this thesis the student has the possibility to do measurements using the HiSPARC detection stations. These detectors are partly maintained by the participating schools. The student has also the opportunity to write a technical thesis discussing maintenance and its problems.

Here we publish a module – for 17 years old students – which we use in the Route-Net to understand the detection of cosmic rays with two scintillator plates containing anthracene. This module is a typical mix of modern physics and chemistry. Also our two years experience in presenting this module to students will be discussed.

**Description of the module anthracene**

1. *Introduction*

This module follows the *de Broglie* module [1]. The detection of cosmic rays in our detection stations situated at the roof of the school occurs with the help of the organic molecule anthracene with molecule formula $C_{14}H_{10}$. The structure of this molecule is seen in Fig. 1.1.

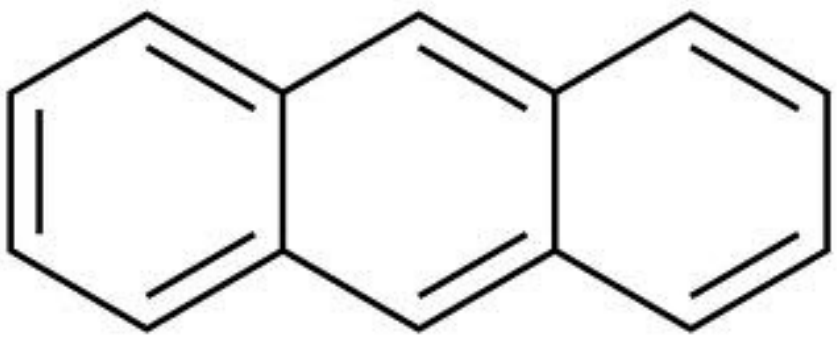

Fig. 1.1. Structure of anthracene.

In this structure C-atoms are situated at the corners of the ring, while H-atoms are not shown. Every straight line joining two atoms represents a chemical bond and consists of two electrons. Actually this picture is not correct because at the carbon-carbon double bonds the electrons are located in between the C atoms. At the chemistry lessons you learned different structures for the aromatic cyclic benzene molecule i.e., the Kekulé structure and the regular hexagon containing a circle in it (Fig. 1.2).

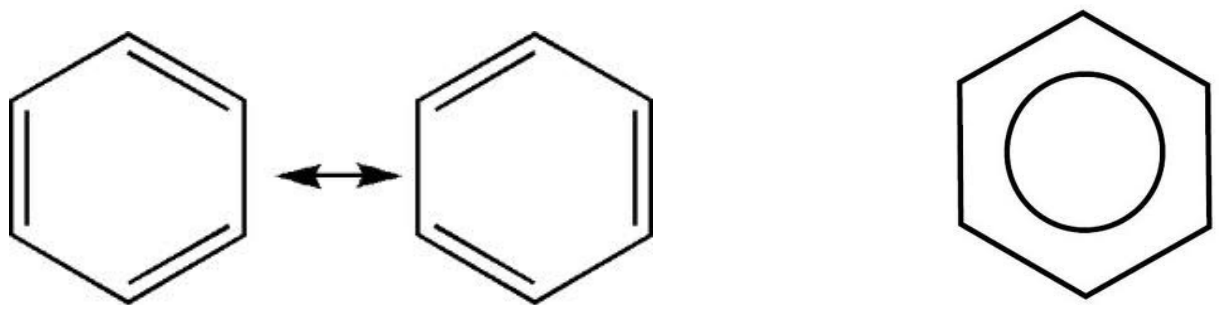

Fig. 1.2. Benzene. Left: the Kekulé structures. Right: modern picture.

The same occurs with anthracene. The two electrons forming the double bond are actually not located between the C atoms, but can move freely over the ring of carbon atoms. Because the C-atom has a covalence of four, the remaining electrons – six in total – move freely and are represented by the circle in the hexagon (Fig. 1.3). This structure was derived from calculations of modern quantum chemistry developed in the last century.

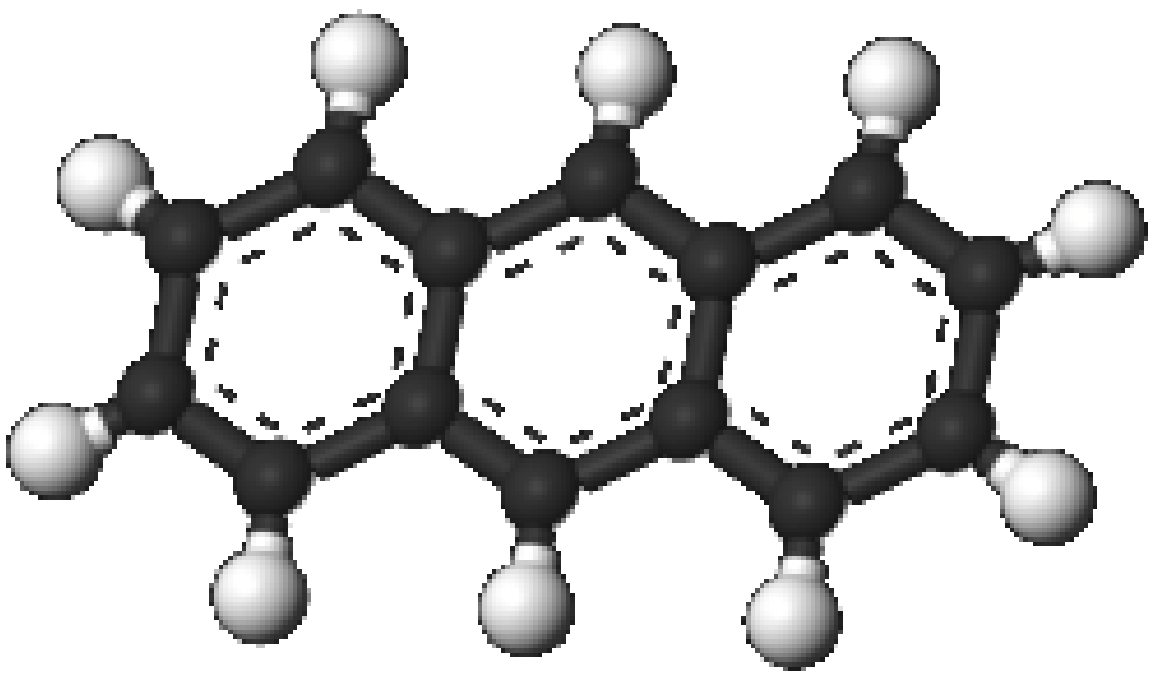

Fig. 1.3. A modern picture of anthracene.

2. *Atomic structure*

To understand how *fluorescence* works in anthracene we will first consider the *structure of the atoms* and the building up of the periodic system of the elements [2]. Then we will treat the *chemical bond* [3,4]. These are necessary steps in understanding *fluorescence*.
The Rutherford description of the atom can be understood as a heavy positively charged nucleus with charge $Ze$ ($Z$ = the ordinal number in the periodic system of the elements – that is the atomic number) about which $Z$ negatively charged electrons rotate.

The following exercises can be solved with the help of *hints* at the end of the module.
Exercise 1. *Before Rutherford, J.J. Thomson had a simpler model of the atom. Describe that model.*
Exercise 2. *Which experiment led to the ideas of Rutherford?*

Later a new, more sophisticated, model of the atom by Bohr was suggested. According to this model the electrons move in orbits around the nucleus, the same as the planets movements in our solar system. With the help of classical mechanics of Newton ($mv^2/r$) and Coulomb's electrostatic law ($Ze^2/r$) , Bohr calculated the different orbits of the electrons i.e., the so-called K-, L-, M- shells. With the idea of different *discrete* orbits, quantum mechanics was introduced in the building up of the atom. The electrons fill the different shells (orbits) according to $2n^2$ with $n$ an integer. Therefore the K-shell can contain 2, the M-shell 8 and the N-shell 18 electrons. In the periodic system this is called a period. Many physical and chemical properties of the atoms can be understood by this system (Fig. 2.1).

| Group → / ↓ Period | 1 | 2 | 3 | 4 | 5 | 6 | 7 | 8 | 9 | 10 | 11 | 12 | 13 | 14 | 15 | 16 | 17 | 18 |
|---|---|---|---|---|---|---|---|---|---|---|---|---|---|---|---|---|---|---|
| 1 | 1 H | | | | | | | | | | | | | | | | | 2 He |
| 2 | 3 Li | 4 Be | | | | | | | | | | | 5 B | 6 C | 7 N | 8 O | 9 F | 10 Ne |
| 3 | 11 Na | 12 Mg | | | | | | | | | | | 13 Al | 14 Si | 15 P | 16 S | 17 Cl | 18 Ar |
| 4 | 19 K | 20 Ca | 21 Sc | 22 Ti | 23 V | 24 Cr | 25 Mn | 26 Fe | 27 Co | 28 Ni | 29 Cu | 30 Zn | 31 Ga | 32 Ge | 33 As | 34 Se | 35 Br | 36 Kr |
| 5 | 37 Rb | 38 Sr | 39 Y | 40 Zr | 41 Nb | 42 Mo | 43 Tc | 44 Ru | 45 Rh | 46 Pd | 47 Ag | 48 Cd | 49 In | 50 Sn | 51 Sb | 52 Te | 53 I | 54 Xe |
| 6 | 55 Cs | 56 Ba | | 72 Hf | 73 Ta | 74 W | 75 Re | 76 Os | 77 Ir | 78 Pt | 79 Au | 80 Hg | 81 Tl | 82 Pb | 83 Bi | 84 Po | 85 At | 86 Rn |
| 7 | 87 Fr | 88 Ra | | 104 Rf | 105 Db | 106 Sg | 107 Bh | 108 Hs | 109 Mt | 110 Ds | 111 Rg | 112 Cn | 113 Uut | 114 Fl | 115 Uup | 116 Lv | 117 Uus | 118 Uuo |

| | | | | | | | | | | | | | | | |
|---|---|---|---|---|---|---|---|---|---|---|---|---|---|---|---|
| Lanthanides | 57 La | 58 Ce | 59 Pr | 60 Nd | 61 Pm | 62 Sm | 63 Eu | 64 Gd | 65 Tb | 66 Dy | 67 Ho | 68 Er | 69 Tm | 70 Yb | 71 Lu |
| Actinides | 89 Ac | 90 Th | 91 Pa | 92 U | 93 Np | 94 Pu | 95 Am | 96 Cm | 97 Bk | 98 Cf | 99 Es | 100 Fm | 101 Md | 102 No | 103 Lr |

Fig. 2.1. The periodic table of the elements.

After the postulate of the duality of matter – particle-wave character – by the Broglie (see the module de Broglie), which was confirmed by later experiments, modern quantum mechanics was introduced.

*Exercise* 3. *Give a synonym for the word postulate.*

Later, with the help of the famous Schrödinger wave equation, one was able to calculate solutions of the H-atom and the energy levels belonging to them. This was an enormous break through in understanding the different properties of the H-atom.

The wave equation is as follows:

$$H\psi = E\psi \qquad (2.1)$$

The Hamiltonian $H$ contains expressions for the movements of the nucleus and electrons and the mutual electrostatic interactions between the electrons, and between the electrons and nucleus. The energies $E$ of the atom can also be found. The solutions of the $2^{nd}$ order differential equation (2.1), the different $\psi$'s, are the so-called wave functions. We will discuss these functions later (especially $\psi^2$).

Exercise 4. *Which interactions will take place between the electrons and the nucleus?*

Exercise 5. *Which expression gives the energy of movement in classical (Newton) mechanics?*

The wave functions (named orbitals) of eq. (2.1) are not simple to be found. For the wave functions the term *orbital* is introduced, because the electrons actually do not move in orbits. The solutions of eq. (2.1) are:

$$1s$$
$$2s,\ 2p_x,\ 2p_y,\ 2p_z$$
$$3s,\ 3p_x,\ 3p_y,\ 3p_z,\ 3d$$
$$\text{etc.} \qquad (2.2)$$

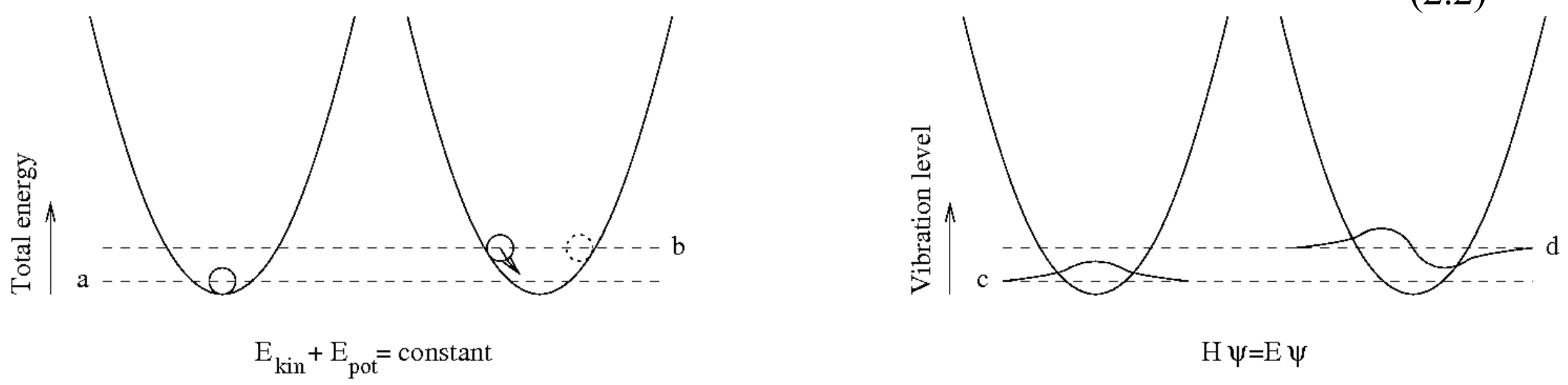

Fig. 2.2. Harmonic oscillator.

Fig. 2.2 shows the energies and wave functions for the ground - (a & c) and first excited state (b & d) of a harmonic oscillator. Examples of harmonic oscillators are the pendulum, swing, mass-spring system and the movement of a marble on a parabolic track. In Fig. 2.2b we notice that the potential energy is at a maximum when the marble is changing direction (see 2.2b). This energy is transformed into kinetic energy, which is at its maximum in the middle of the parabola (see 2.2a).

Exercise 6. *If we look at an arbitrary moment at Fig. 2.2b, explain where there is the highest chance to find the particle (classical approach).*

The diagrams in 2.2c and 2.2d give the solutions of the 2nd order differential equation $H\psi = E\psi$. The chance or probability of finding the particle can be calculated with $\psi^2$ (quantum mechanics – Born). As shown before we speak of a chance to find the electron and therefore we cannot speak anymore of an *orbit* in an atom. The chance to find an electron somewhere around the nucleus is not limited to a certain orbit anymore, but the nucleus is surrounded by an electron cloud (*orbital*).

## 3. *Spin*

In experiments with Ag-atoms (or H-atoms) in an inhomogeneous magnetic field it was shown that the electron can take only two positions in space (Stern-Gerlach experiment). This can be understood as if the electron possesses two discrete angular momentums called up and down spin. This is completely different  from what is expected by the prediction of classical mechanics. Because the electron has an electrical charge, the spinning is accompanied by a small magnetic field (Fig. 3.1).

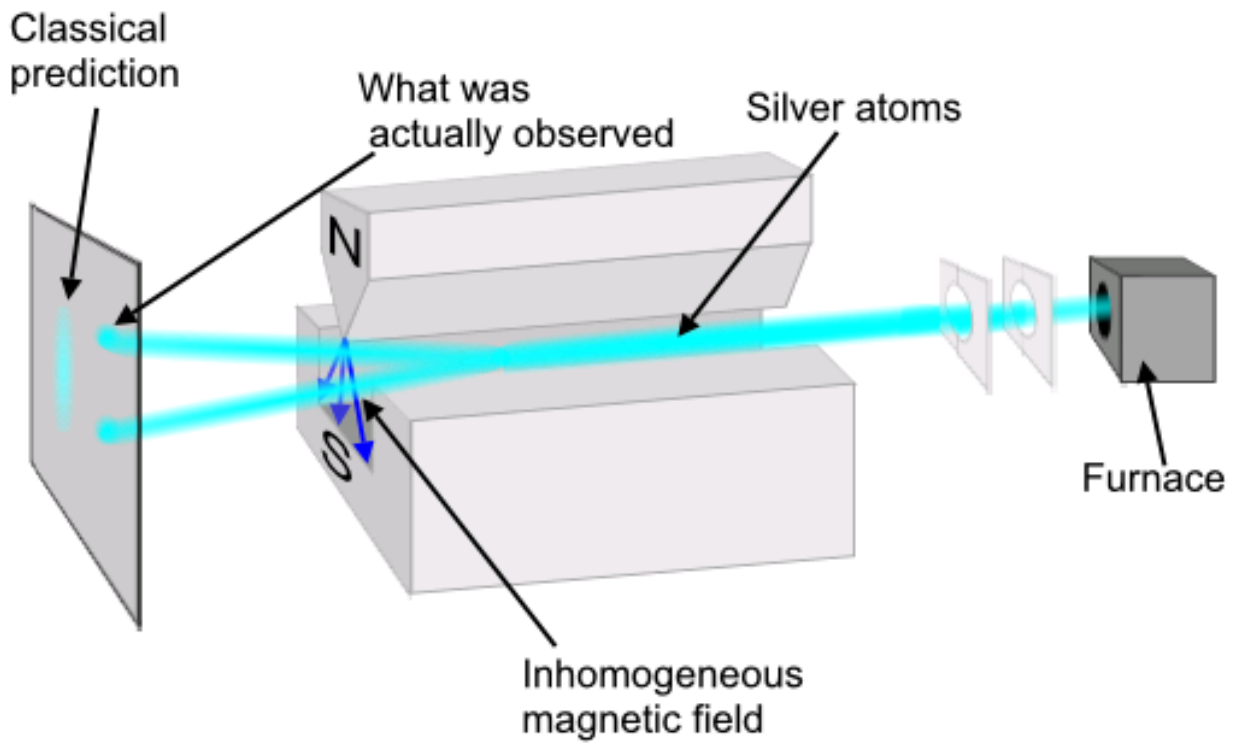

Fig. 3.1. Electron spin.

Exercise 7. *Explain the letters* N *and* S *in fig. 3.1. Which directions will the spin take with respect to the magnetic field lines?*

To understand the “aufbau” (German for building up) of the periodic table Pauli introduced the rule that no more than two electrons may occupy any given orbital and if two do occupy one orbital then their spins must be paired with opposite spin angular moments $S = \pm\frac{1}{2}\left(\frac{h}{2\pi}\right)$. This is called the exclusion principle. The symbol $h$ is Planck’s constant (the value you can find in a Table book).
Because of this exclusion principle, the first 1*s* shell contains a maximum of 2 electrons. The second shell contains a maximum of 8 electrons divided over the 2*s*, $2p_x$, $2p_y$, $2p_z$ orbitals (see Fig. 2.1).
Let us have a closer look at the H-atom: H has atom number 1. In the ground state the electron occupies the 1*s* orbital. Fig. 3.2 shows the shape of the different orbitals and Fig. 3.3 the energy levels belonging to them.

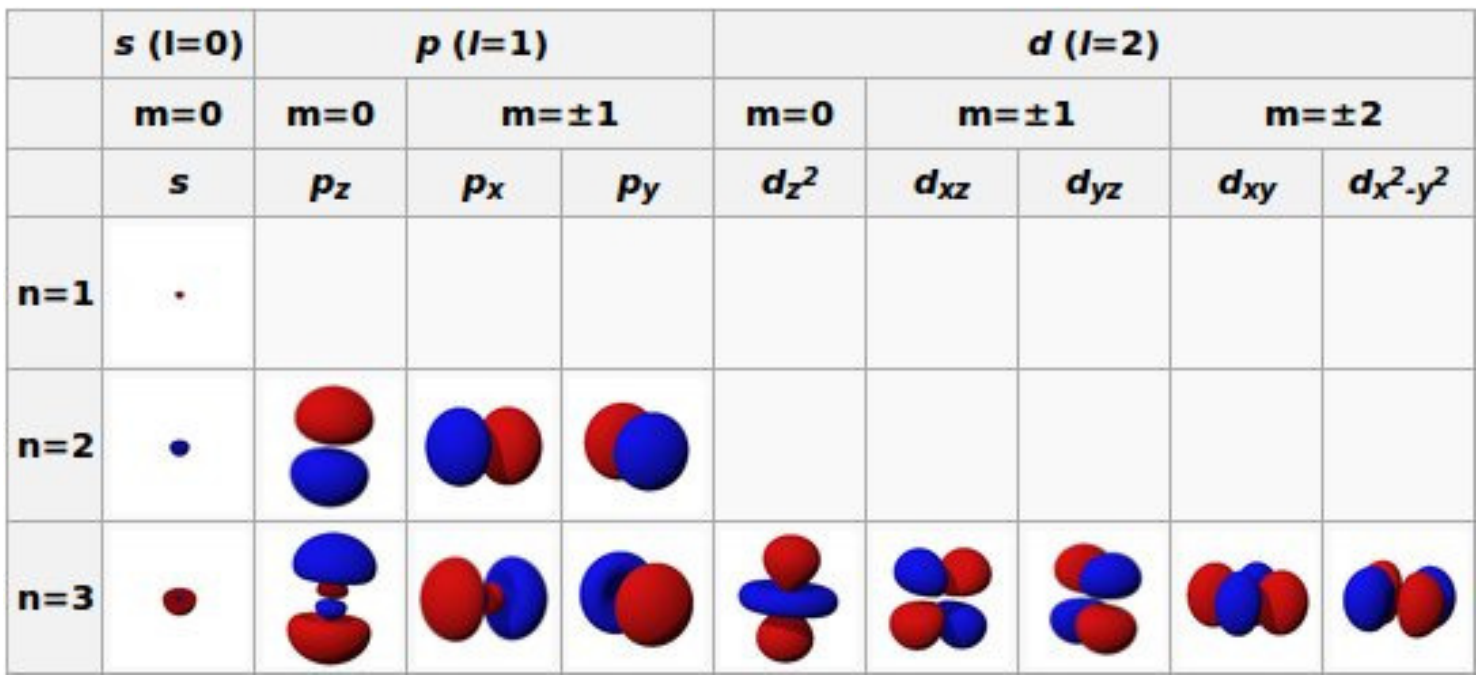

Fig. 3.2. Orbitals.

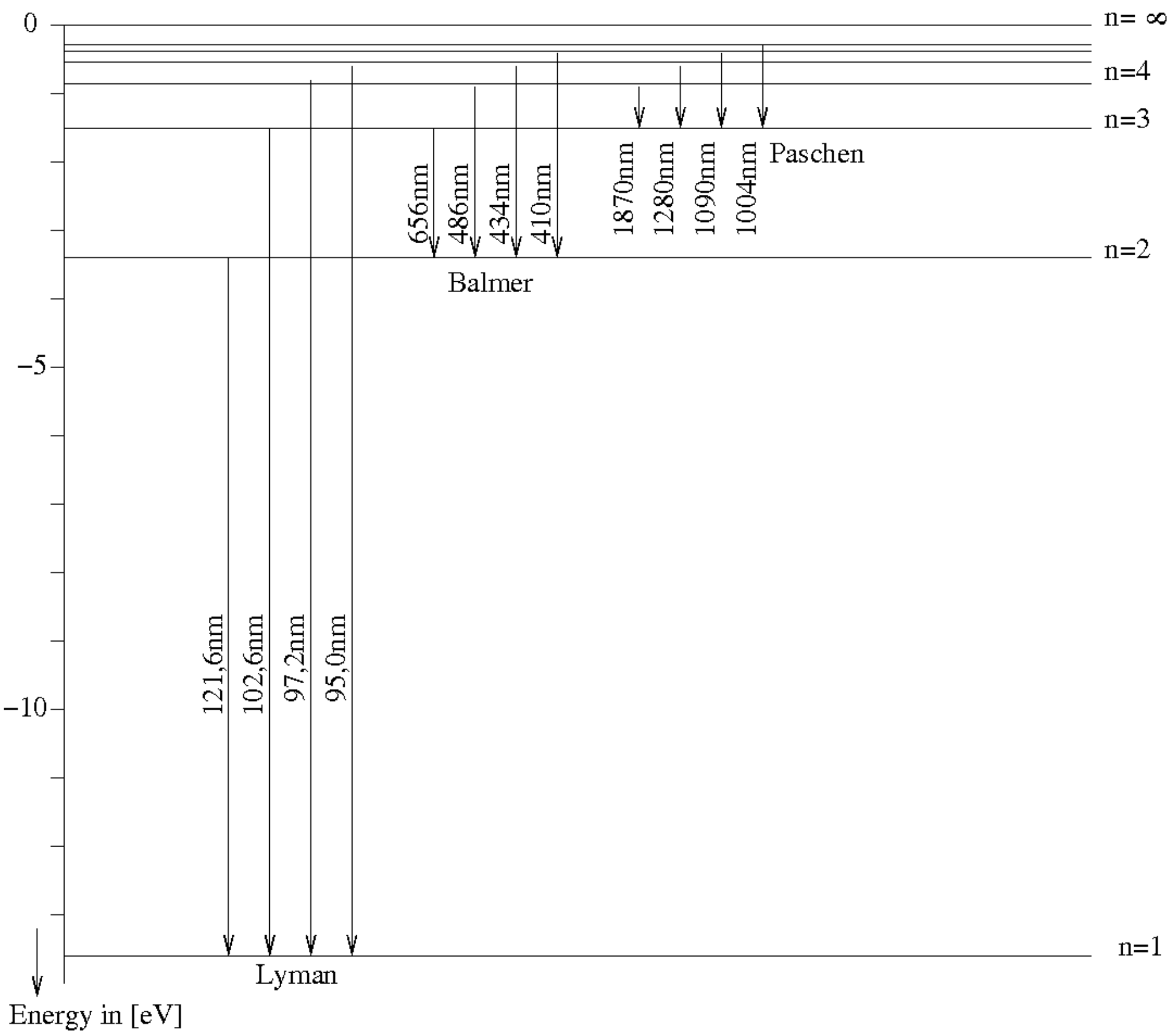

Fig. 3.3. Emission stemming from the H-atom.

The atom can absorb energy in the form of radiation (light). Emission occurs when the atom loses energy in the form of radiation. The arrows in Fig. 3.3 show emission in the H-atom.

## 4. *Singlet and triplet*

After treating the H-atom we will have a look at the He atom. Now we have to deal with two electrons. According to the exclusion principle these electrons occupy in the ground state the lowest 1*s* orbital with opposite spins. A new phenomenon in quantum physics is the property of electrons that they are *indistinguishable.* Therefore we cannot label the electrons and do not know the position of the electrons [4].
The mathematical expression is then as follows:

$$N\{1s(1)\alpha(1)1s(2)\beta(2) - 1s(1)\beta(1)1s(2)\alpha(2)\} = N\{1s(1)1s(2)[\alpha(1)\beta(2) - \beta(1)\alpha(2)]\} \quad (4.1)$$

*N* is a so-called normalisation factor, which we will not treat further. The function $1s(1)$ means electron 1 occupies orbital 1*s*, α means spin up and β spin-down.

*Exercise* 8. *Explain in your own words the meaning of the term* $1s(1)\alpha(1)1s(2)\beta(2)$.

In expression (4.1) we notice that each electron with α spin is *paired* with an electron with β spin. This paired state is called a *singlet.*

*Exercise* 9. *What does this mean for the total magnetic moment?*

We will now consider the He-atom in its first excited state. The electrons can occupy the 1*s* as well as the 2*s* orbital. From now on we only look at the spin part of expression (4.1), otherwise it will be to complicated. Expressions (4.2) give the wave functions of the first excited state of the He-atom.

$$
\begin{gathered}
N\{\alpha(1)\alpha(2)\} \\
N\{\beta(1)\beta(2)\} \\
N\{\alpha(1)\beta(2)+\beta(1)\alpha(2)\} \\
N\{\alpha(1)\beta(2)-\beta(1)\alpha(2)\}
\end{gathered}
\tag{4.2}
$$

*Exercise* 10. *Give the total magnetic moment of the four expressions of equation (4.2).*

The first three expressions of eq. (4.2) form a so-called *triplet* state (magnetic) and the last one a *singlet* (not magnetic). In the *appendix* we will explain why the first three functions belong together. Fig. 4.1 shows the energy diagram of the He-atom. The left part is formed by singlet states (parahelium) and the right part by triplet states (orthohelium).

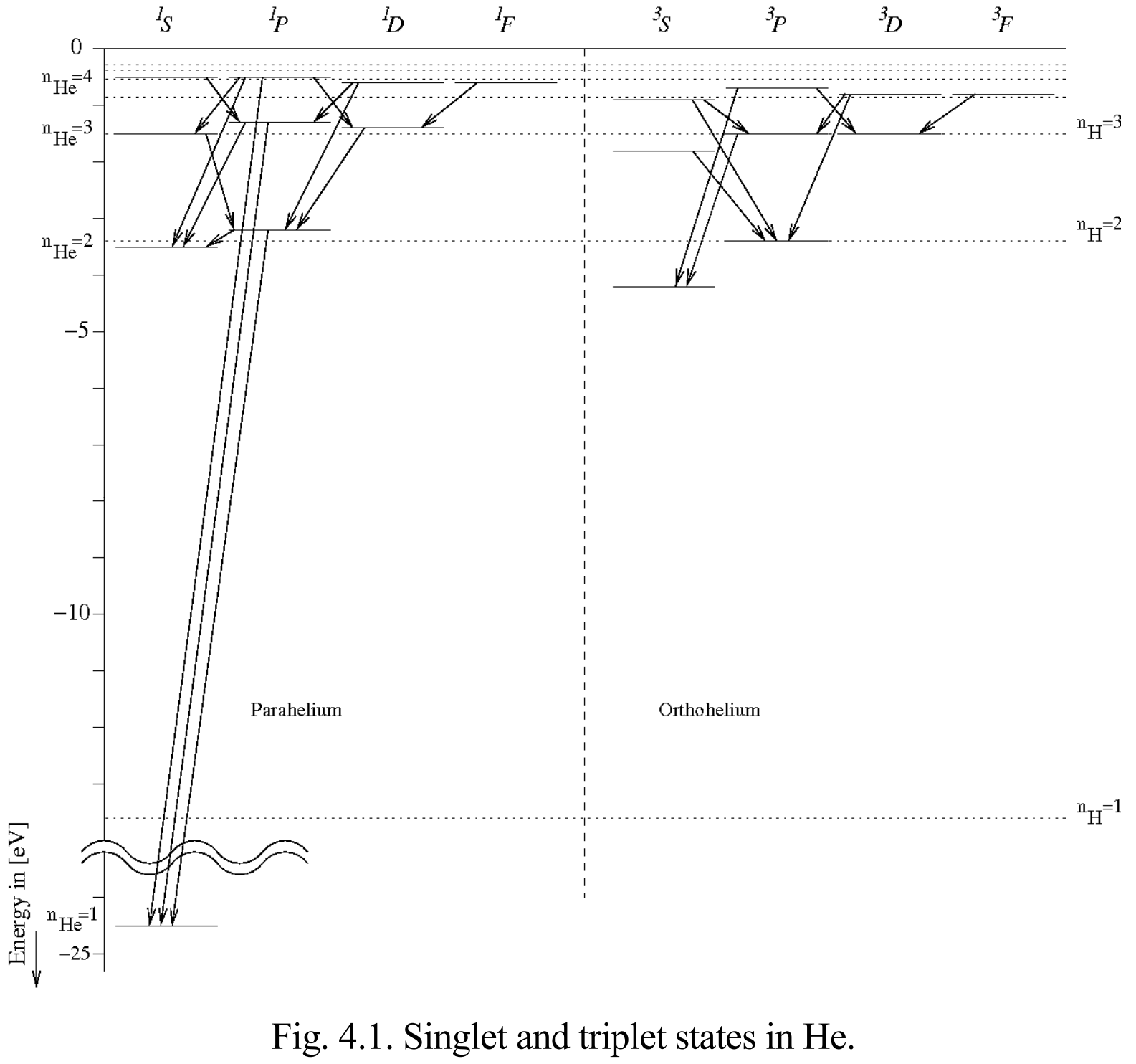

Fig. 4.1. Singlet and triplet states in He.

*5. Molecular bond in the hydrogen molecule*

The question arises how can we understand the chemical bond in the $H_2$ molecule [3]. Formerly we solved the problem with the Lewis octet rule (inert gas configuration), but now we will start with a quantum chemical approach.

The 1*s* orbital of the first atom H (atom *a*) will combine with the 1*s* of the second atom H (atom *b*).

$$\Phi_{\pm} = N\{1s(a) \pm 1s(b)\} \tag{5.1}$$

Here the + as well as the – (linear) combinations are possible. The + combination is the molecular bonding orbital, the – combination is the molecular anti-bonding orbital. The anti-bonding orbital lies higher in energy. Every $H_2$ molecule possesses two electrons (see the He-atom), therefore the bonding orbital can accommodate a maximum of 2 electrons, spin-up and spin-down, according to the exclusion principle (Fig. 5.1). In the excited states again singlets and triplets will occur, just as in the He-atom.

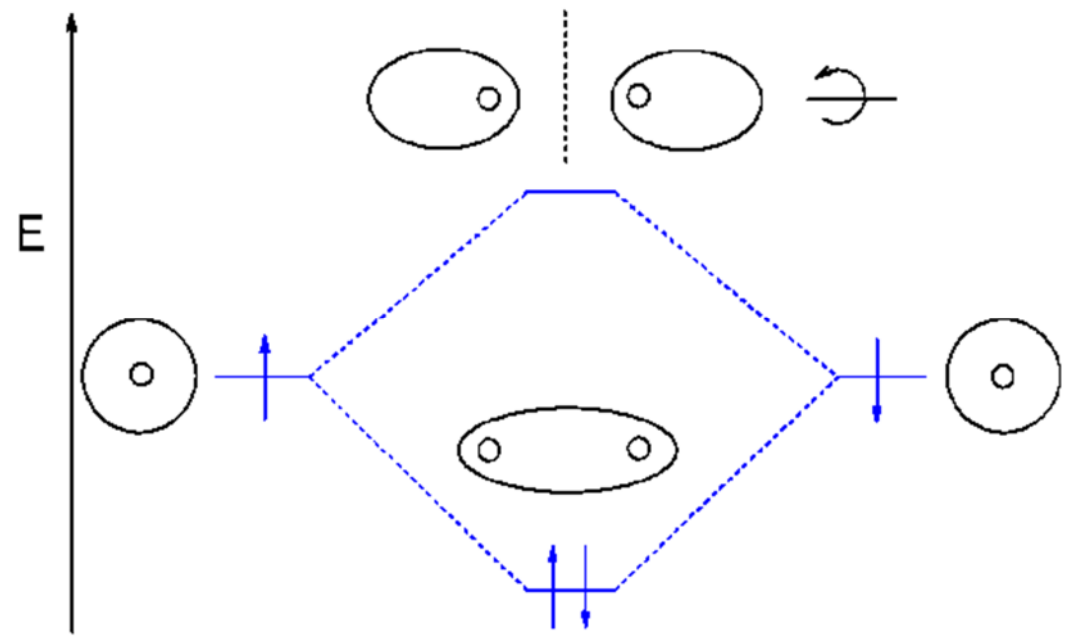

Fig. 5.1. Singlet ground state in the $H_2$ molecule The circles are the 1*s* states of the H-atom. They form bonding and anti-bonding molecular orbitals in $H_2$.

*Exercise* 11. *Draw in Fig. 5.1 the first excited state and write down the spin wave functions belonging to it.*

In contrast to atoms molecules can vibrate. This gives rise to extra energy levels (molecules can vibrate faster or slower). In Fig. 5.2 the horizontal axis gives the distance between the H-atoms, the vertical axis the energy. Therefore there is a favourable distance between the atoms with the strongest bonding. The figure partly looks like a parabola and shows different energies levels, which a di-atomic molecule can assume. The energies drawn into the parabola represent the different vibration energies.

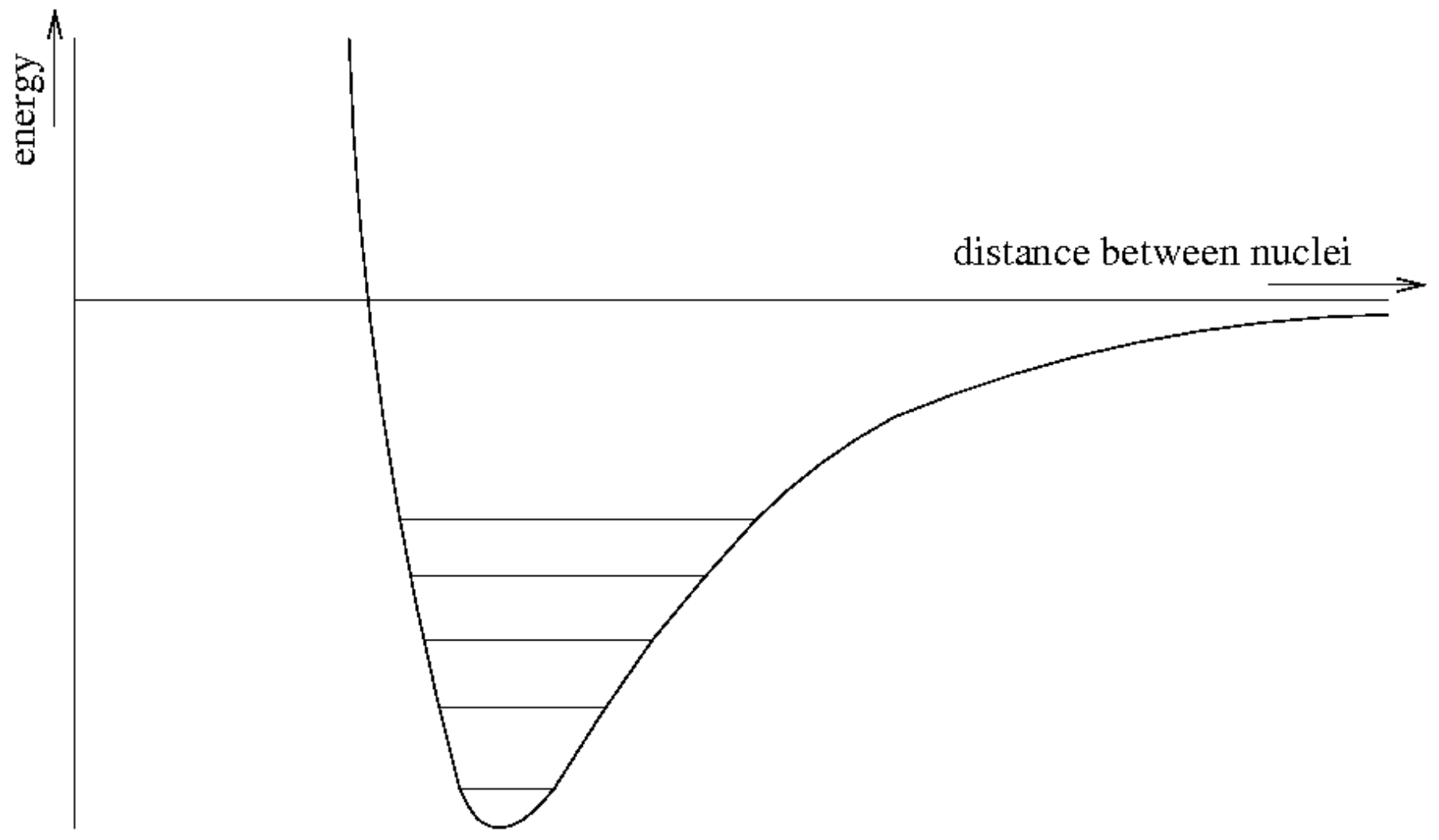

Fig. 5.2. Energy levels of an $H_2$ molecule. The drawn lines are vibrational states.

6. *Bonding in antracene.*

The atomic number of carbon 6. Two electrons are present in the 1*s* orbital and four electrons occupy the 2*s*, $2p_x$, $2p_y$, $2p_z$ orbitals. Now each of these orbitals possesses 1 electron.

*Exercise* 12. *Do you have any idea why there is 1 electron in each orbital and there is no pairing allowed by Pauli?*

Looking at the antracene molecule we have bonding between the C-atoms and bonding between the C- and H-atoms. We do not take into consideration the low lying 1*s* orbital of the C-atom, because this orbital does not contribute to the chemical bond. We only take into account the 1*s* orbital of H and the 2*s*, $2p_{x,y,z}$ of C.
The C-H bonds are formed by the 1*s* of the H-atom and the $2p_x$ of the C-atom. The C-C bonds are formed by the $2p_y$ orbitals of the C-atoms. These bonding orbitals are lying in the plane of the molecule. Chemists call them σ-bonds. The $2p_z$ orbitals form so-called π-bonds. These are perpendicular to the plane of the molecule.

*Exercise* 13. *Draw the* $p_z$ *orbitals in antracene.*

*Fluorescence* and *phosphorescence* in antracene are stemming only from the π-electrons. We now consider the two outer π-electrons. The other electrons form pairs according to Pauli. These two electrons can be treated in the same way as the two electrons in the He-atom i.e., a singlet ground state and an excited singlet and triplet state are formed.
Fluorescence is only possible between an excited singlet state and a singlet ground state. The transition occurs in $10^{-8}$ second. Phosphorescence is actually a 'forbidden' transition and therefore lasts ‘much’ longer. It occurs between a triplet state and the singlet ground state. If energy is absorbed (as will happen when a cosmic ray hits the molecule) the molecule is excited and falls back via the vibrational energy levels and will then emit a light photon. The molecule is again in its ground state.

*Exercise* 14. *Explain figure 6.1.*

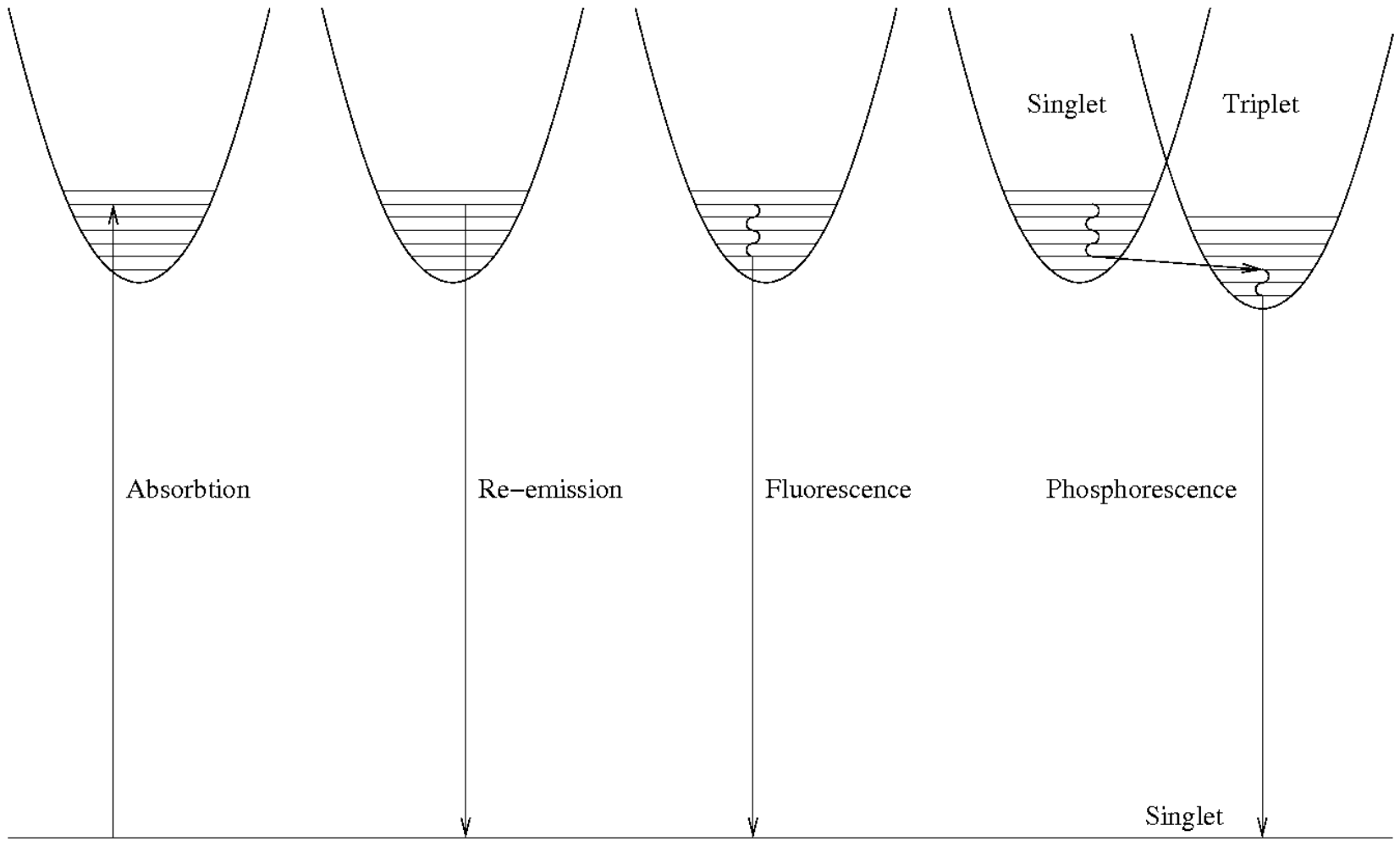

Fig. 6.1. Absorption and decay in a di-atomic molecule.

The energy diagram of Fig. 6.2 gives possible energy transitions in antracene [4].

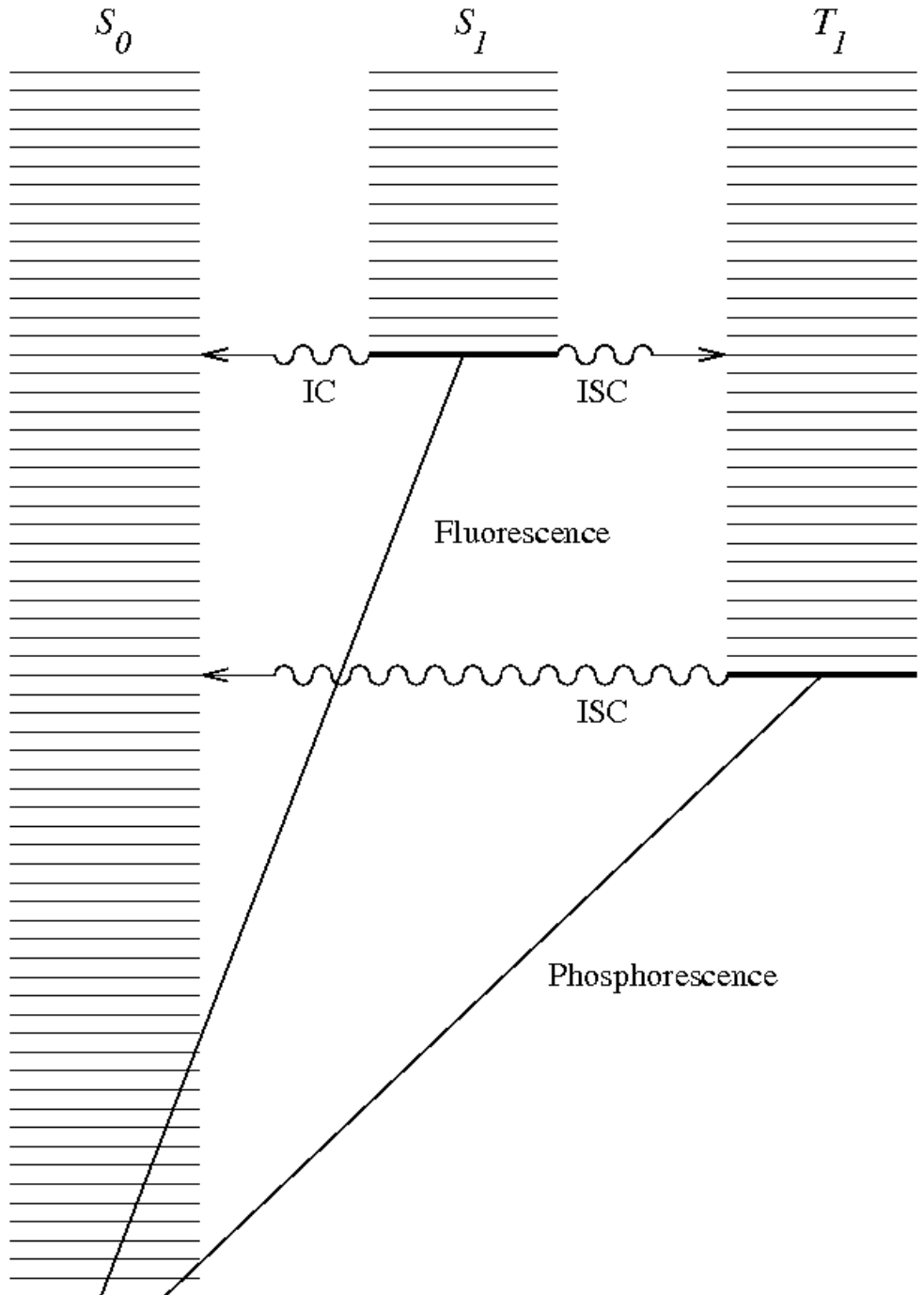

Fig. 6.2. Decay in antracene.

Now we will give an explanation of the different symbols in Fig. 6.2. ISC means intersystem crossing. These are transitions from excited singlet states to vibration levels of the triplet state. IC means internal conversion, transitions of an excited singlet state to vibrations of the singlet ground state.

*Exercise* 15. *What do you observe if you look at the vibration levels of $S_0$ (singlet ground state) and $S_1$ (first excited singlet state)?*

Fluorescence takes place between the excited singlet level ($\pi^*$) and the singlet ground state ($\pi$). This gives rise to a photon in the visible light spectrum. This photon is then multiplied by a photo-multiplier.

*Exercise* 16. *Why do you think that we use the symbols $\pi$ and $\pi^*$?*

*Appendix*

Another way of formulating the exclusion principle is the property of the total wave function that by an exchange of the labels of the electrons the wave function must be anti-symmetric. The singlet ground state in He will be [4]:

$$\Psi(1,2) = \{1s(1)\alpha(1)1s(2)\beta(2) - 1s(1)\beta(1)1s(2)\alpha(2)\} = \{1s(1)1s(2)[\alpha(1)\beta(2) - \beta(1)\alpha(2)]\}$$

Now we will exchange the labels of the electrons:

$$\Psi(2,1) = \{1s(2)1s(1)[\alpha(2)\beta(1) - \beta(2)\alpha(1)]\} = \{1s(1)1s(2) - [-\alpha(2)\beta(1) + \beta(2)\alpha(1)]\} = -\Psi(1,2).$$

The first part in the expression is the symmetric orbital function and the second part is the anti-symmetric spin function. This is a singlet state leading to a particular energy.
As you can see:

$$\alpha(1)\alpha(2),\ \beta(1)\beta(2) \text{ and } \alpha(1)\beta(2) + \beta(1)\alpha(2)$$

are symmetric spin functions. To make the total function anti-symmetric we have to multiply these functions with an anti-symmetric orbital part.
As shown above: $\alpha(1)\beta(2) - \beta(1)\alpha(2)$ is an anti-symmetric spin function leading to a triplet state with lower energy.

*Hints*

The following hints give a tool to solve the exercises.

*Exercise* 1. Which particle was discovered by Thomson in 1897? Give the charge of that particle. How is this charge compensated? Describe now the Thomson atom.

*Exercise* 2. Collision experiments with α-particles on atoms (gold foil). Give the charge of the α-particle. What was the outcome of these experiments. Describe now the Rutherford atom.

*Exercise* 3. Synonym for postulate. See (shorter) Oxford English dictionary.

*Exercise* 4. Consider electrostatic attraction and repulsion between charges (Coulomb's law).

*Exercise* 5. See your physics books.

*Exercise* 6. Where does the particle remain longest? Consider two situations: a. Where is the fastest movement of the particle (marble)?; b. Where is there no movement? Give now an answer to the exercise.

*Exercise* 7. How are the poles of a magnet called? Realise that the spin of an electron is magnetic and possesses only *two* directions in a magnetic field (quantum mechanics). Also *N* – *N* as well as *S* - *S* is repulsive and *N* – *S* is attractive.

*Exercise* 8. 1*s*(1) – particle 1 occupies orbital 1*s*; α(1) – particle 1 possesses an α spin, etcetera.

*Exercise* 9. Remember α spin = +½ and β spin = -½. Calculate now the sum.

*Exercise* 10. Again α spin = +½ and β spin = -½. Calculate now the sum in the four expressions.

*Exercise* 11. Put one electron in the low lying molecular (MO) orbital 1*s*(a) + 1*s*(b) and the other electron in the higher lying MO 1*s*(a) – 1*s*(b). Look also at the solutions of the He-atom. Give now the answer.

*Exercise* 12. a. Put two electrons together in one orbital (this is allowed by the Pauli principle). b. Put one electron in one orbital and the other electron in another orbital. Compare the energies of these two situations with regard to the electron repulsion (Coulomb's law).

*Exercise* 13. The $p_z$ orbital is perpendicular to the plane of the antracene molecule.

*Exercise* 14. From left to right.

- Absorption of energy (a light photon) from the singlet ground state to a vibrational level of the singlet excited state.
- Emission from an excited vibrational singlet state to the singlet ground state (in the form of light with the same frequency as in the left figure).
- First step: Emission from an excited singlet vibrational level to the vibrational ground state of an excited singlet state. Second step: Emission from an excited singlet state to the singlet ground state This gives rise to *fluorescence* and has a different frequency.
- First step: Emission from an excited singlet vibrational level to an excited vibrational level of a triplet state. Second step: Emission from an excited vibrational level of the triplet state to the ground level of the triplet state. Third step: Emission from the triplet state to a the singlet ground state. This gives rise to *phosphorescence*. This process takes more time.

*Exercise* 15. Overlap of vibrational levels from a singlet state and a triplet state.
*Exercise* 16. The $p_z$ orbitals form the π electronic states, which give rise to singlets and triplets.

**Discussion**

In the school year 2010-2011 six students followed the course at our school, which was then given for the first time. Different modules were treated including fluorescence. The students were enthusiastic and the modules gave them an idea of modern physics. They did not find it easy to follow but it gave them a flavour of what is happening in the detectors.

In the year 2011-2012 ten students followed the course. It looked like that this group understood this module better than the first group. One of their tasks was telling younger students about the contents of this and other modules. This was quite a success. We can recommend this approach.

**Curriculum**

Niek Schulheiss is teacher in physics, computing and science at the Zaanlands Lyceum, located in Zaandam, the Netherlands. Beside teaching, he manages four HiSPARC detection stations and develops course-ware for secondary school students. Since 2009 he also developed Route-Net and jSparc for HiSPARC at Nikhef, Amsterdam, with a FOM-LIO appointment. Now he works at the possibilities of doing measurements using the complete HiSPARC detection grid.

Dr. Tom Kool is a also a secondary school teacher in physics and chemistry. He worked at the Zaanlands Lyceum and is still active in research at the University of Amsterdam. He wrote different peer reviewed scientific articles about condensed matter materials. With 1987 Nobel prize winner Alex Müller he wrote a book about oxide perovkites [5]. He also gives courses in quantum mechanics, chemical bond and thermodynamics at university level.